# High-fidelity achromatic metalens imaging via deep neural network

**Yunxi Dong**[1,*], **Bowen Zheng**[1,*#], **Hang Li**[1], **Hong Tang**[1], **Yi Huang**[1], **Sensong An**[1], **Hualiang Zhang**[1,#]

[1]University of Massachusetts Lowell, Lowell, 01854 MA, USA
[#]bowen_zheng2@uml.edu, hualiang_zhang@uml.edu

**ABSTRACT**

Meta-optics are attracting intensive interest as alternatives to traditional optical systems comprising multiple lenses and diffractive elements. Among applications, single metalens imaging is highly attractive due to the potential for achieving significant size reduction and simplified design. However, single metalenses exhibit severe chromatic aberration arising from material dispersion and the nature of singlet optics, making them unsuitable for full-color imaging requiring achromatic performance. In this work, we propose and validate a deep learning-based single metalens imaging system to overcome chromatic aberration in varied scenarios. The developed deep learning networks computationally reconstruct raw imaging captures through reliably refocusing red, green and blue channels to eliminate chromatic aberration and enhance resolution without altering the metalens hardware. The networks demonstrate consistent enhancement across different aperture sizes and focusing distances. Images outside the training set and real-world photos were also successfully reconstructed. Our approach provides a new means to achieve achromatic metalenses without complex engineering, enabling practical and simplified implementation to overcome inherent limitations of meta-optics.

## 1. Introduction

The advancement of modern camera systems has led to multi-element lens configurations to minimize optical aberrations and achieve high-resolution imaging. However, these systems sacrifice compactness. Metasurfaces, the two-dimensional metamaterial analog of optical components, provide transformative opportunities to realize high-performance optics within substantially reduced volumes. Here, we utilize metalenses - metasurface lenses with carefully engineered nanoscale scattering elements that impart precise phase profiles - to demonstrate imaging capabilities analogous to conventional refractive optics. Notably, metalenses overcome the challenge of spherical aberration that has persisted in refractive optics. By imparting precise phase delays with subwavelength spatial resolution, they facilitate diffraction-limited focusing absent from traditional refractive optical systems due to the spherical shape of traditional lenses. Additionally, the capability to readily adapt the phase profile through computational nanophotonic design of the meta-atoms grants flexibility and customizability surpassing conventional optics.

However, a pivotal roadblock for the wide deployment of meta-optics is chromatic aberration. Due to significant material dispersion and dispersive responses of metasurfaces, different spectral components passing through metalenses will focus on disparate spatial planes, negatively impacting image quality. Existing strategies to mitigate chromatic aberration include cascaded multi-layer metalenses(*1–4*), interleaving meta-atoms for different wavelengths(*5–7*), metalens arrays(*8*), dispersion correction phase mask(*9–12*), increased focusing depth(*13*) and computational optimization and correction of phase profiles(*14*, *15*). But these approaches increase system complexity while sacrificing other performance metrics such as scalable high-yield fabrication, imaging quality and freedom of material choices. Consequently, a single meta-lens solution capable of full-color aberration-free imaging under diverse operating conditions remains elusive.

In this paper, we successfully demonstrate correction of chromatic aberration to achieve an achromatic metalens camera through integration of a custom-designed metalens with a commercial imaging sensor, coupled with deep learning algorithms. Our deep learning-based computational imaging approach refocuses and restores missing information of the raw captured images for RGB channels, effectively converting a single chromatic metalens camera into an achromatic imaging system. With this strategy, light (i.e. broadband optical signals) can be manipulated within substantially thinner flat optical components compared to the state-of-the-art while still maintaining full-color and aberration-free operation. To collect multi-spectral training data, we employ a 3D-printed adapter for the integration of metalens onto a commercial camera. As for the computational imaging backend, a universal deep neural network architecture built on U-net is used to achieve direct chromatic aberration correction. By training with raw images under varying conditions, the model reliably enhances image quality, removes chromatic aberration, effectively reconstructs the photos either from or outside of the training dataset. The trained model can also be used for enhancing real-world captures, which further demonstrates its capability to replace complex lens assemblies for high-quality full-color imaging.

## 2. Results

### 2.1 Imaging system workflow, DL model and experimental setups

An achromatic single metalens imaging system presents considerable difficulty due to the requisite restoration of all color channels lacking ideal imaging responses. To address this issue, we integrate deep learning networks as the computational backend to directly enhance the chromatic responses of the raw image captures. A highly automated workflow for collecting and pre-processing raw images was developed to enable the proposed deep learning approach as depicted in Fig. 1. Specifically, Fig. 1a shows the optical path with the metalens directly mounted on a camera, where $d$ denotes the object distance and $A$ denotes the aperture diameter. The aperture is placed in front of the metalens L, and its diameter is equal to or smaller than the metalens to block light outside the metalens area. Fig. 1b illustrates the assembled metalens with a 3D-printed mount on a commercial camera (Sony Alpha a7R IV). Changing the object distance *d* and aperture diameter *A* alters working conditions of the metalens, making it suitable for a variety of applications. One of our goals is to devise universal deep learning models as illustrated in Fig. 1d to accommodate different combinations of *d* and *A*.

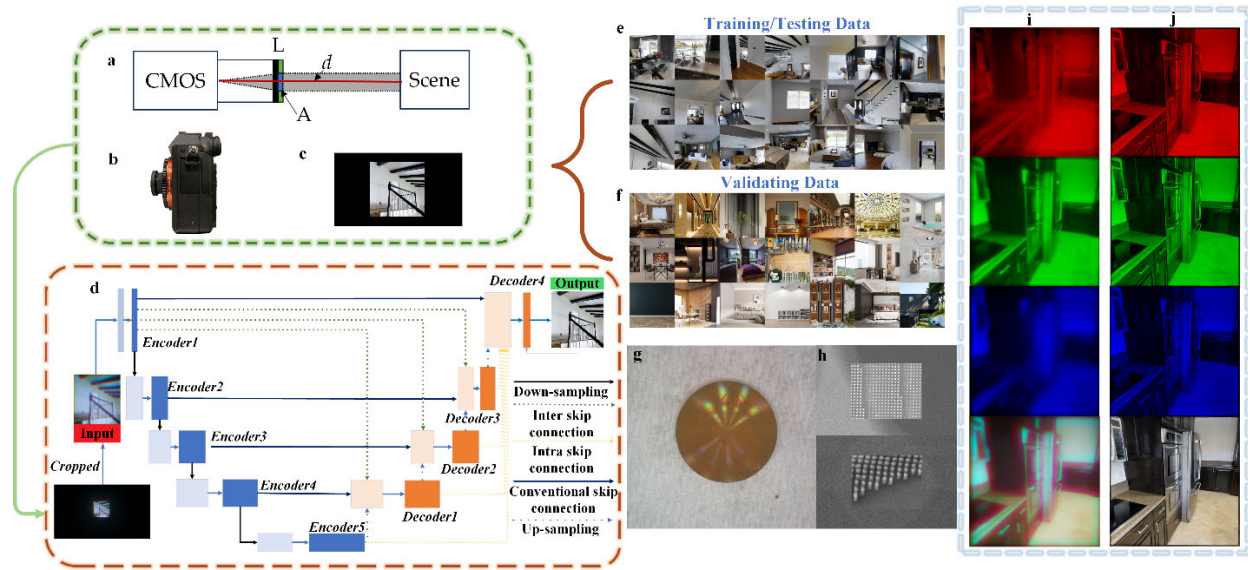

**Figure 1. Overview of metalens imaging and reconstruction.** (a) Optical path in metalens camera system, with light passing through aperture, then metalens which directly focuses light onto CMOS image sensor. (b) Photograph of fabricated metalens mounted on a commercial camera. (c) Example source image displayed on monitor. (d) Schematic of image reconstruction workflow, including preprocessing of raw images and deep learning model for reconstruction. (e) Sample images from training and testing datasets used for deep learning model. (f) Additional validation image samples showing different objects and color representations. (g) Photograph of fabricated metasurface lens. (h) Scanning electron microscope (SEM) images showing nanostructured meta-atoms comprising metalens. (i) Cropped regions of raw red, green, and blue color channel subimages directly captured by metalens camera. (j) Reconstructed red, green, and blue channel subimages after processing through the proposed deep learning network.

In this work, we utilized monitors of various sizes to display images (as depicted in Fig. 1c), as well as accommodating different object distances (*d*). To conform to the image circle of the metalens, the images' aspect ratios were intentionally set to 1, with all remaining monitor pixels set to black. The resulting captured image, positioned at the sensor's center as shown in the left corner of Fig. 1d, was cropped to eliminate black pixels and used as input for the developed deep learning network.

Inspired by the successful application of image super-resolution networks, we developed a U-Net-structured deep learning model. This state-of-the-art architecture, widely applied in image processing tasks(*16*, *17*), features skip connections that bridge contracting and expanding paths, enabling the capture of both global and local contexts. Our model enhances the original U-Net architecture by incorporating multiple skip and residual connections between layers, capturing multi-scale contexts and providing nuanced features as shown in Fig. 1d. Inter-skip connections link the encoder and decoder blocks within the U-Net model, while intra-skip connections, exclusive to the decoder blocks, link different layers within them, and conventional skip connections denote the original connections within the U-Net model(*18*) The structure of the encoder and decoder blocks, comprising several convolutional and upsampling layers, is detailed in the supplementary material.

To train the deep learning models, we utilized the Taskonomy indoor scene dataset(*19*), examples of which are shown in Fig. 1e. This dataset contains 1024 × 1024-pixel images from various buildings, providing diversity in environments and objects under consistent lighting. The resolution matched our 1920 × 1200-pixel monitors used for data collection, as illustrated in Fig. 1c. For each combination of object distance *d* and aperture diameter *A*, we selected 1000 images from the dataset to display on the monitors and capture with our metalens camera system. Of the 1000 raw images, 800 were used for training and 200 for validation for each setting (with different *d* and *A* combination).

After convergence of the network training process, we applied an additional validation set with completely different objects, colors, and lighting conditions to assess the performance of the trained network, as shown in Fig. 1f. The results were consistent across both the training/testing sets and the additional validation set. One of the examples of these results is shown in Fig. 1i and Fig. 1j, which features raw captures using a 4 mm aperture diameter to capture a scene at a 50 cm distance, and its reconstructed counterpart using the trained network. The raw images predominantly contain sharp image components in the green channel, with the red and green channels significantly out of focus, aligning with our assumptions for the employed metalens. Remarkably (as shown in Fig. 1j), processing the raw captures through the deep learning network yielded a reconstructed image with clear images across RGB channels, indicating the network's ability to eliminate achromatic aberration by refocusing the image on its all three channels. Each channel benefits from an improvement in sharpness, and achromatic full-color imaging is achieved by combining all three channels. For further performance analysis details, please refer to the supplementary material.

Our proposed deep learning engine for computational achromatic metalens imaging presents several notable advantages. Firstly, the incorporation of deep learning renders further metalens design for chromatic aberration correction unnecessary (which leads to simplified metalens implementation and reduced cost). Secondly, it eliminates the requirement for supplementary devices or steps from the initial photo capture to the final image reconstruction. Lastly, this method can be readily implemented on any commercial or scientific optical systems. To the best of our knowledge, the proposed image reconstruction network represents the first successful application of a deep learning tool for addressing aberrations in chromatic meta-lens imaging captured directly from a commercial camera.

## 2.2 Designed metalens and integration

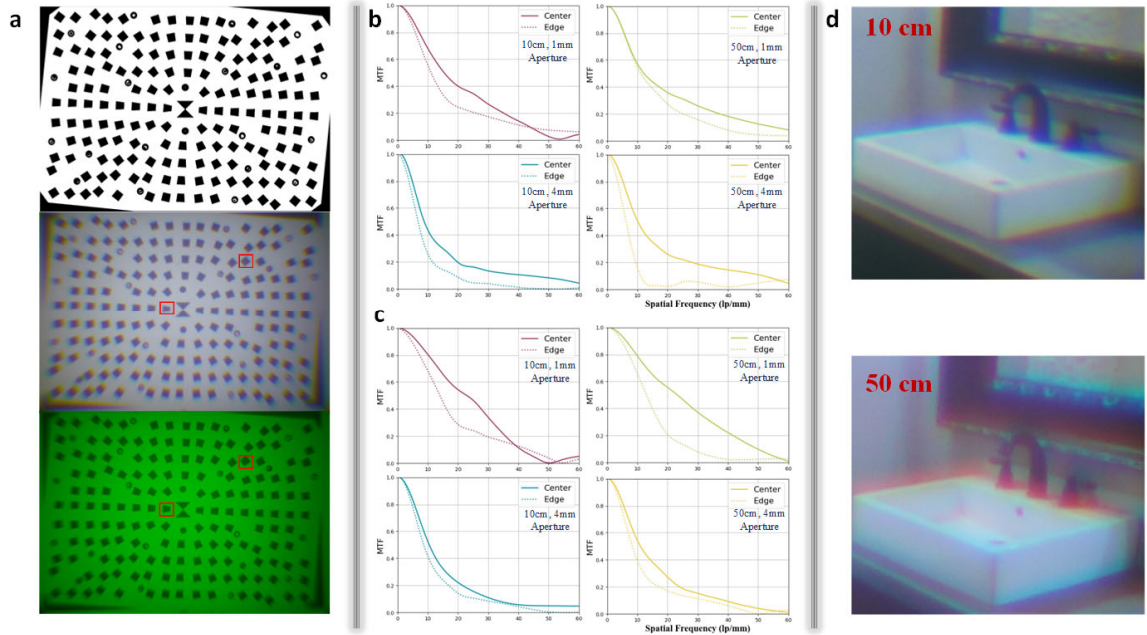

**Figure 2. Metalens performance characterization.** (a) Test charts imaged under white and green illumination. (b-c) Modulation transfer functions (MTFs) for the center and edge regions of images captured under white and green light. Four combinations of object distance and aperture diameter were tested for each condition.

In this work, a hyperbolic phase profile(20) was employed for the metalens, which offers several advantages. Firstly, the hyperbolic phase profile works well with the external aperture, as the entire lens area is designed to focus to the geometric center point. This allows for the inclusion of an external aperture without altering the focal length or compromising the imaging uniformity. Additionally, misalignment between the aperture and the metalens does not impact the imaging performance, as long as the transparent part of the substrate is fully blocked. Notably, the aperture size plays a crucial role as it impacts both the imaging resolution and chromatic aberration. Smaller apertures improve resolution and reduce chromatic aberration image-wide, yet larger apertures enable greater light transmission beneficial for low-light conditions. Our approach provides the flexibility to incorporate various sizes of external apertures using different 3D-printed holders, eliminating the need to fabricate metalenses of different sizes. Furthermore, it opens up the possibility of integrating a mechanical leaf aperture, similar to those found in traditional lenses. Our metalens was designed and fabricated on a 10 mm by 10 mm Silicon-on-Sapphire wafer with 230 nm Silicon thickness. It has a 5 mm diameter with 7 mm focal length, and the meta-atoms were optimized for operation at a wavelength of 526 nm. More information about the metalens can be found in the supplementary material.

Meanwhile, hyperbolic phase profile has notable drawbacks, including compromised peripheral image quality stemming from unoptimized edges. This is manifested as reduced sharpness and increased chromatic aberration towards the image boundaries. For example, it is oberved that edge trapezoids in Fig. 2a appear less defined versus the center. Rainbow effects under white light further underscore greater chromatic aberration at the periphery. Additional limitations of hyperbolic lenses arise from variable field-of-depth and lateral chromatic aberration across different focal planes and object distances. This leads to captured images exhibiting differently sized in-focus areas and distinct chromatic aberration patterns depending on distance, as evidenced by the Modulation Transfer Function (MTF) results in Fig. 2(b-c)(21).

Fig. 2(b) and 2(c) display four combinations of 10 cm (representing close focusing) and 50 cm focusing distances (simulating focusing to infinity) with 1 mm and 4 mm aperture diameters (f-numbers of 7 and 1.75). Fig. 2(b) shows center and edge MTF curves derived from denoted trapezoids in Fig. 2(a) under white backlight on monitors. Contrastingly, Fig. 2(c) exhibits the photo's green channel with green backlight. Regardless of conditions, a notable MTF difference exists between center and edges, with higher center values. Overall, it is clear that increasing aperture diameter decreases MTF values significantly, especially at edges, thus a universal deep learning network should be trained on various aperture sizes to handle these dramatic differences.

Chromatic aberration reduces MTF values, as is evident from comparing the center of smaller apertures to the edge of larger ones between Fig. 2(b) and 2(c). The marked MTF difference between green channel and white light data suggests chromatic aberration primarily causes decreased image quality in these scenarios. Conversely, for instances involving the center of larger apertures and the edge of smaller ones, the difference becomes less pronounced, with both white and green MTF values displaying similar trends. This suggests that specific image reconstruction algorithms should be applied for enhancing small and large aperture cases, given the unique sources of image blurriness in each.

Although the MTF curves do not exhibit significant differences across varying object distances, the patterns of color fringing do display noticeable variations. Fig. 2(d) presents two images, cropped from the center of photos captured at 10 cm and 50 cm distances. No apparent differences in sharpness exist between these two images, yet the color fringing patterns around white edges differ significantly. The 10 cm photo exhibits blue to cyan and yellow to orange color fringing transitions at the far and near ends of the faucet, respectively. However, a reverse fringing transition pattern is observed when images were captured at 50 cm (it exhibits orange to yellow and cyan to blue transitions instead). These distinct color fringing patterns underscore the influence of object distance on the effects of chromatic aberration in captured images.

The MTF curves and color fringing analyses reveal key insights into factors impacting image quality in meta-optics systems. Specifically, aperture size and object distance significantly influence aberrations and resolution. Therefore, to comprehensively improve photo quality, the effects of varying aperture diameter and shooting distance must be considered in tandem. In our work, we have been focusing on studying these two factors and their interactions, to determine optimal deep learning model architectures, formation of proper experiment setup and training strategies to enhance image quality across diverse operating conditions. In general, we could train universal deep learning models applicable to a wide range of potential use cases rather than being constrained to a narrow set of parameters.

## 2.3  Imaging results

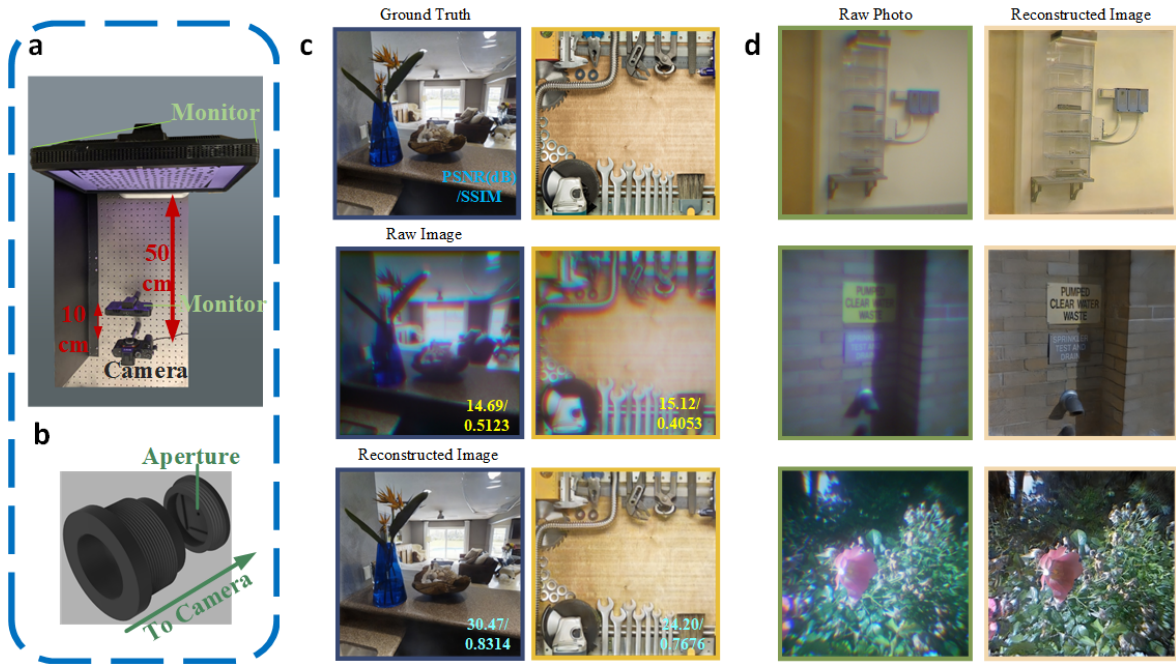

**Figure 3. Experimental setup and image reconstruction examples.** (a) Top view of the photo capturing setup. (b) 3D-printed metalens holder with adjustable aperture. (c) Reconstructed images from testing data (first column) and validation set (second column). (d) Real-world photos captured through the metalens with different aperture sizes and reconstructed by the proposed network.

The setup for raw photo capturing is depicted in Fig. 3a. It consists of two monitors of varying sizes fixed on an optical table, facing towards a commercial camera integrated with a single metalens. The larger monitor (24" HP LP2475W) and the smaller one (5.2" Atomos Ninja V) are located 50 cm and 10 cm away from the camera, respectively. The monitors were employed individually during the experiment. The camera, set on a post, was adjusted to be parallel with the monitor and captured images using the center of the CMOS sensor, with the raw image displayed in Fig. 1d. A 3D-printed holder, depicted in Fig. 3b, was designed to attach the metalens to the camera. This holder is composed of two parts: the upper component is threaded into a C-Mount adapter attached to the camera, while the lower section holds the 10 mm by 10 mm metalens sample. These two parts are threaded together, with the aperture located on the lower component to facilitate the interchange of varying aperture sizes. Upon assembly, the metalens is pushed into place by the upper component, reducing any undesired gap between the lens and the holder.

Based on the previously described setup, raw images were captured and used to train the proposed deep learning models. The model's performance, as demonstrated in Fig. 3c, was assessed with test images from both the training and validation sets. The ground truth image, randomly chosen from Taskonomy dataset, the raw image directly sourced from the camera, and the reconstructed image from the output of the deep learning model, are all presented in Fig. 3c. To quantitatively measure the enhancement in image quality from raw to reconstructed images, we utilized two primary metrics: the peak signal-to-noise ratio (PSNR)(*22*) and the structural similarity index measure (SSIM)(*23*). Higher PSNR values typically indicate reduced noise and improved image detail fidelity, while SSIM indicates measure similarity between two images. By comparing the PSNR and SSIM values of both the raw and reconstructed images to the ground truth image, we were able to quantify image quality improvements enabled by the proposed technique.

The PSNR and SSIM values, calculated for the respective raw and reconstructed images, are displayed in Fig. 3c (shown at bottom right). It is obvious that our deep learning models effectively mitigated chromatic aberrations and increased image sharpness by refocusing all color channels. The model's reconstruction process successfully restored accurate color representations and considerably enhanced overall image contrast, yielding a gain of over 10 dB in PSNR and a 35% increase in SSIM values for the training set images. The computations revealed notable enhancements in image quality through our reconstruction method compared to the raw images.

To fully validate our model's versatility, we conducted extensive testing on entirely new types of images beyond the indoor training data. As shown in Fig. 1f, we utilized an additional validation set of diverse scenes with various objects, lighting conditions and tones. Without any further training or parameter tuning, our pre-trained model successfully reconstructed these never-before-seen images. As evidenced in Fig. 3c (right column), our network reliably restored color and focus for these general validation images. Quantitatively, the developed deep learning network enhanced image quality by over 9dB in peak signal-to-noise ratio and approximately 36% in structural similarity index. These impressive gains aligned with those observed on the indoor training images, conclusively demonstrating the model's robustness and applicability to real-world scenes.

Lastly, we applied the model to reconstruct real-world scenes taken both indoors and outdoors. Unlike controlled scenes using monitors, real-world objects feature a significantly larger depth-of-field and varied lighting conditions, making reconstruction remarkably more challenging. Despite these complexities, the network consistently performed well. The raw and reconstructed images are shown in Fig. 3d, featuring one indoor and two outdoor photos. In the raw images, reduced dynamic range (evidenced by hazing) and chromatic aberration are noticeable, which diminish image quality and make distinguishing objects and characters difficult, especially in ample light conditions and near high-contrast areas. It is noted that, after the deep learning model reconstructed the images, the image quality improved significantly under all conditions, regardless of ambient lighting or shooting distances. These results further validate our model's universality and adaptability beyond its training set.

## 3. Discussion

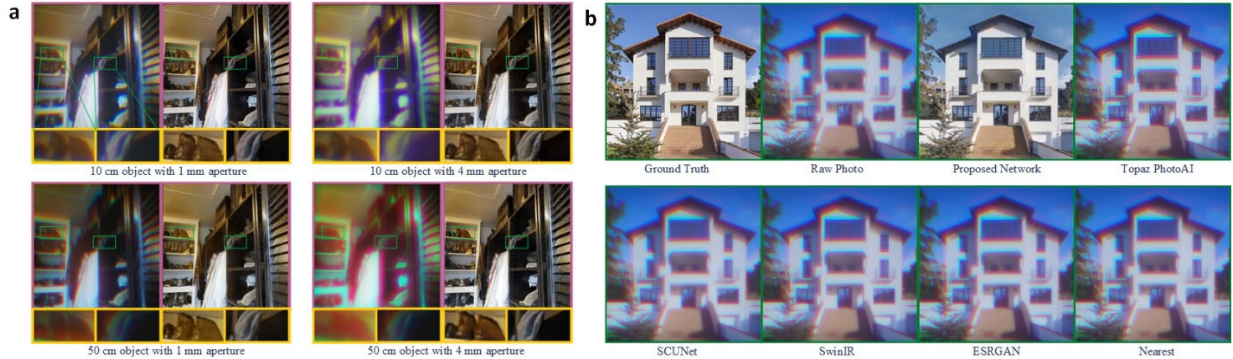

**Figure 4. Image reconstruction for different scenarios and algorithms.** (a) Raw and reconstructed images for all experimental combinations, with enlarged details from image centers and edges below. (b) Ground truth, raw single metalens image, and reconstructions from the proposed network and other existing networks. Only the proposed network successfully reconstructs the single metalens image.

As demonstrated in previous sections, the proposed deep learning approach successfully restored full-color images from raw captures of the single metalens camera. Both quantitative metrics and visual interpretation confirmed significant enhancement of image quality compared to the unprocessed raw images exhibiting chromatic aberration. This represents a major advancement for single metalens imaging systems (which have faced persistent challenges in achieving achromatic performance).

Inherent material dispersion limits metalens bandwidth when relying solely on optical and metasurface design innovations. Despite efforts exploring multi-layer systems, new materials, and hybrid meta-refractive concepts, realizing wide-band achromatic responses from a single nanostructured meta-optics device has remained elusive. Our proposed deep learning-based computational imaging engine provides a transformative solution to overcoming these physical constraints. By applying specialized deep learning models directly to raw captured images, we accomplish full-color aberration-free imaging without requiring complex metalens/metasurface engineering, reducing design and fabrication difficulties, improving tolerance, and enabling faster turnaround.

To further validate performance and gain additional insights, we conducted detailed studies on the reconstructed images. Fig. 4(a) shows further analysis and comparisons across setups. A high dynamic range image from the training set was chosen given the challenge of preserving both dark and bright details using metalenses (this is evident in the raw photos where even using a small aperture leads to hazing and chromatic aberration. It is also observable through the zoomed-in view of details at the center and edge of the image and placed on the bottom of each photo). Enlarging the aperture rapidly worsens image quality (e.g. making the shoes at the edge of the photos barely distinguishable). The central image quality exceeds the edges but still lacks details in dark regions. Increasing object distance also degrades edge image quality, as at constant angular resolution and chromatic aberration ratio, greater distances lead to reduced resolution and increased chromatic aberration per pixel. Notably, our deep learning models can handle the varying challenges across different setups and consistently produce promising results. As shown in Fig. 4(a), the reconstructed images on the right of each setup remove strong color fringing and accurately restore dark region details without losing the bright details. The reconstructed images' strong similarities across different setups indicate the developed deep learning networks are universal and insensitive to aperture size, lighting conditions, and object distance.

**Table 1.** Comparison of quantitative measure of imaging performance from different algorithms in Fig. 4(b).

|  | Raw Photo | Proposed network | Topaz PhotoAI | SCUNet | SwinIR | ESRGAN | Nearest |
|---|---|---|---|---|---|---|---|
| PSNR | 14.568 | 21.447 | 14.554 | 14.527 | 14.565 | 14.544 | 14.568 |
| SSIM | 0.532 | 0.791 | 0.561 | 0.515 | 0.537 | 0.531 | 0.532 |

To demonstrate the uniqueness of our reconstruction approach, it is necessary to show that existing general-purpose computational imaging networks fail to effectively reconstruct images from our metalens. As shown in Fig. 4b, we benchmarked leading super-resolution and enhancement models by upsampling our raw images and then downsampling the outputs to 512 × 512 pixels for comparison(*24–26*). To validate generalization ability, the test image is chosen from the validation set that the network was never trained on, and nearest neighbor image scaling method (labeled as Nearest in Table 1) is used as control to validate the up- and down-sampling process. It is clear that none of the existing networks can remove chromatic aberration of metalenses. While some enhanced local details, they failed to improve global color and focus. Quantitative PSNR and SSIM analyses were conducted, and the results are listed in Table 1. It is concluded that our specialized deep learning network significantly outperformed these existing methods designed for generic imagery. This confirms the necessity of tailoring the model to the unique artifacts and distortions in raw metalens images. The customized network architecture and training process are essential to learn the intricacies of correcting meta-optics aberrations computationally.